\documentclass{article}
\usepackage{spconf,amsmath,amssymb,graphicx,booktabs,multirow,xcolor}
\usepackage[hidelinks]{hyperref}
\hypersetup{pdftitle={ARIS: Low-Resource Glass-Box Neural Source-Filter Synthesis for Phonetic Stimulus Manipulation},pdfauthor={Yiran Ding, Wenwei Xu},pdfsubject={ICASSP 2027},pdfkeywords={speech manipulation, phonetic stimuli, source-filter synthesis, formant control, differentiable DSP}}
\graphicspath{{./}{./figures/}}
\title{ARIS: Low-Resource Glass-Box Neural Source--Filter Synthesis\\for Phonetic Stimulus Manipulation}
\name{Yiran Ding, Wenwei Xu}
\address{Leiden University Centre for Linguistics (LUCL), Leiden University, The Netherlands}

\begin{document}
\ninept
\maketitle

\begin{abstract}
Phoneticians often need to construct stimuli in which specific acoustic cues are precisely manipulated while preserving decent speech quality. Classical synthesis and modern neural methods sit along a trade-off between precise parametric control and high fidelity, and neural synthesis typically demands more data than phoneticians can easily obtain. We present ARIS (Analytic Resonant Interpretable Synthesis), a neural source--filter model that pairs neural parameter estimation with deterministic DSP synthesis. Every control is a coefficient of the synthesizer, so $F_0$, formants and the glottal source can be edited directly. On five small single-speaker corpora in three languages, ARIS resynthesizes speech with quality comparable to WORLD and edits single parameters more accurately than Praat KlattGrid, with negligible crosstalk between cues. Compared with HiFi-Glot, pre-trained on a large corpus and fine-tuned on the same data, ARIS scores slightly lower on predicted naturalness but reproduces the recordings more faithfully and manipulates them more precisely. Audio samples: \url{https://n1r.github.io/ARIS_nsf/}.
\end{abstract}


\begin{figure*}[t]
  \centering
  \includegraphics[width=0.9\textwidth]{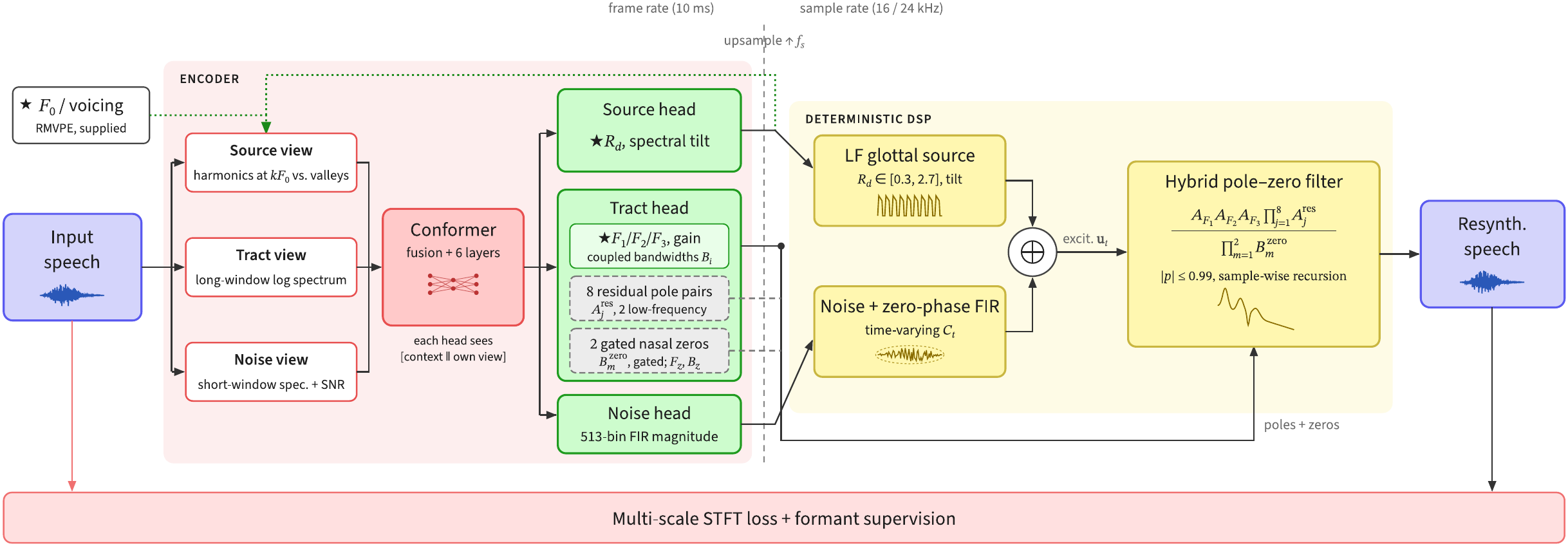}
  \caption{ARIS architecture. The encoder fuses source, tract and noise observations with a Conformer, with one output head per synthesizer component; $F_0$ and voicing come from RMVPE. The vocal tract cascades explicit $F_1$--$F_3$ resonators, 8 residual pole pairs and 2 gated zeros ($\star$: editable; dashed: learned but not exposed).}
  \label{fig:arch}
\end{figure*}

\section{Introduction}
\label{sec:intro}

Manipulated speech is a basic instrument of phonetic research: by changing one or two acoustic cues and keeping everything else comparable, experimenters reveal how listeners weight those cues. Classic findings on categorical perception~\cite{liberman1957boundaries}, vowel and consonant identification~\cite{lisker1970voicing,hillenbrand1999resynthesized}, lexical tone~\cite{blicher1990mandarintone} and voice quality~\cite{klatt1990voicequality} were obtained this way, and neurolinguistic studies use such stimuli~\cite{baek2026dynamic,normanhaignere2025temporal} to probe cortical encoding of speech. Malisz et al.~\cite{malisz2019modern} discuss three aspects of synthesis for this purpose. The stimuli must be \emph{natural}, because unnatural speech alters how listeners process it; \emph{controllable}, so that an edit can be specified and verified in units such as Hz or semitones, both locally within an utterance and globally across a continuum; and producible with modest \emph{resources}, because phonetic studies rarely have the data and computation that modern synthesis relies on.

Classic tools trade one requirement for another. Formant synthesizers such as KlattGrid~\cite{klatt,weenink2009klattgrid} give full parametric control but struggle to reproduce natural recordings. Analysis--resynthesis methods such as STRAIGHT~\cite{kawahara1999straight} and WORLD~\cite{world} stay closer to the recording, but their control is limited: WORLD, for instance, represents the vocal tract as a smooth spectral envelope without explicit formants. LPC-based resynthesis, as in Winn's formant-continuum script~\cite{winn2019continuum}, filters the LPC residual through editable formants. The residual, however, still carries noise and spectral structure that the low-order LPC model does not capture, and these leak into the output as artifacts.

Neural vocoders~\cite{hifigan} can produce high-fidelity speech, but their intermediate representations, whether mel spectrograms, codec features or other learned latents, entangle many acoustic attributes, so fine control is possible but hard to apply reliably~\cite{li2025continuum,lameris2025voicequalityvc}. Differentiable DSP (DDSP) instead builds the synthesizer from signal-processing components: NSF~\cite{nsf} and DDSP~\cite{ddsp} drive source--filter and harmonic-plus-noise models with neural networks. SawSing~\cite{sawsing} replaced the source with a sawtooth signal, which enables training with little data and computation, and GOLF~\cite{golf,yu2024golf} combined glottal-flow wavetables with differentiable all-pole filters, which also lets it model the phase of the voice, and was later made efficient for sample-wise time-varying filtering~\cite{torchlpc}. A second strand, closely linked to phonetic studies, exposes phonetically meaningful controls within neural synthesis. Source--filter vocoders already achieve flexible $F_0$ control~\cite{yoneyama2023usfgan,yoneyama2022sifigan}. Wavebender GAN~\cite{wavebender} further predicts mel spectrograms directly from a small set of core parameters and synthesizes them with HiFi-GAN, and HiFi-Glot~\cite{hifiglot} conditions differentiable resonant filters on a neural glottal excitation. In all cases, the potential bottleneck is the control-to-signal mapping: it has to be learned from data, so it holds only where the training data cover the control space. Wavebender GAN was trained on 24\,h of LJ~Speech with pitch augmentation, yet manipulated $F_1$ and $F_2$ less reliably than $F_0$. HiFi-Glot extended the training data to 1664\,h. Phonetic research often cannot meet such data requirements (e.g., for low-resource languages, dialects or specific speech phenomena), and manipulations such as shifting pitch or formants beyond their natural range may easily fall outside the data distribution.

We build on this line of work and propose ARIS (Analytic Resonant Interpretable Synthesis), designed for phonetic stimulus manipulation. ARIS is a \emph{glass box}: every control is a coefficient of a deterministic synthesizer. $F_0$ drives the glottal source, $R_d$ shapes the glottal pulse, and each of $F_1$--$F_3$ sets the pole pair of its own resonator. The network only learns the analysis from speech to DSP parameters, so a single speaker's recordings suffice. Our contributions are threefold: (1)~a neural source--filter synthesizer with explicit formant resonators, residual poles and gated nasal zeros that, trained on under an hour of one speaker, reaches copy-synthesis quality comparable to WORLD on five corpora in three languages and reproduces recordings more faithfully than a fine-tuned HiFi-Glot; (2)~precise single-cue phonetic edits, with about half the $F_2$/$F_3$ error of KlattGrid and negligible crosstalk between cues; (3)~an open-source toolkit for phonetic researchers, with examples of global scaling, of vowel, tone and intonation continua, and of local edits within utterances.

\section{Method}
\label{sec:method}

ARIS is an analysis--synthesis model (Fig.~\ref{fig:arch}): an encoder estimates frame-level controls from the waveform, with $F_0$ and voicing taken from RMVPE~\cite{wei2023rmvpe}, and a deterministic decoder converts these controls back to speech. To build a stimulus, the user edits the controls globally or locally, as a ratio or in physical units, and resynthesizes.

\subsection{Encoder}
Given the small amount of training data, we give the encoder observations that are as informative and accurate as possible, one per decoder component. The \emph{tract} branch receives a 64\,ms log-power spectrum, and the \emph{noise} branch a 16\,ms spectrum with harmonic and noise envelopes. The \emph{source} branch, following GOLF~\cite{golf}, contrasts harmonics at $kF_0$ with valleys at $(k{-}\tfrac12)F_0$ to isolate the glottal source. A 6-layer Conformer (256\,dim, 8 heads) fuses them, and one output head per decoder component maps the result, every 10\,ms, to bounded parameters: $R_d$, source tilt, frame gain, $F_i$ and bandwidths $B_i$, residual-pole and zero parameters, and noise-filter magnitudes.

\subsection{Decoder}
The decoder has no learnable parameters. The periodic excitation $e_h$ is read from a Liljencrants--Fant (LF) glottal-flow-derivative wavetable~\cite{lf} indexed by $R_d\in[0.3,2.7]$, sampled at the phase from $F_0$ and gated by voicing. The aperiodic excitation $e_\nu$ is Gaussian noise shaped by a time-varying zero-phase FIR filter. The output is $y=g\,h*(e_h+e_\nu)$, with frame gain $g$ and vocal-tract filter $h$. Unlike GOLF, ARIS factors this filter into interpretable sections:
\begin{equation}
H(z)=\frac{\prod_{m=1}^{2}B_m(z)}{\prod_{i=1}^{3}A_{F_i}(z)\,\prod_{j=1}^{8}A^{\mathrm{res}}_{j}(z)},
\end{equation}
where each formant section is $A_{F_i}(z)=1-2r_i\cos\theta_i\,z^{-1}+r_i^2z^{-2}$ with $\theta_i=2\pi F_i/f_s$ and $r_i=e^{-\pi B_i/f_s}$, and pole radii are clipped at 0.99 for stability. The explicit formants range over $F_1\in[220,1100]$, $F_2\in[700,3400]$ and $F_3\in[2200,4800]$\,Hz, while 8 residual pole pairs absorb the spectral detail beyond $F_1$--$F_3$ and 2 switchable gated zeros model nasal antiresonances (22 poles in total). Editing $F_i$ thus changes only its own section.

\subsection{Training}
The model is trained end to end with $\mathcal{L}=\mathcal{L}_{\mathrm{MSS}}+\lambda_F\mathcal{L}_F+\lambda_P\mathcal{L}_P$ ($\lambda_F{=}2$, $\lambda_P{=}1$), where $\mathcal{L}_{\mathrm{MSS}}$ is a multi-resolution STFT loss at four resolutions~\cite{schwaer2023multiscale} and $\mathcal{L}_P$ a periodicity loss over 64 bands. The formant loss $\mathcal{L}_F$ supervises $F_1$--$F_3$ and their bandwidths with Praat Burg estimates on voiced frames (smooth-$\ell_1$), so formant targets are needed only for training. Each model trains for 50k steps with Adam (lr $2{\times}10^{-4}$, batch size 16, 2\,s segments). With 9.1\,M parameters and 2.1\,GB peak memory, training fits on a single consumer GPU (we used an RTX 4060 and a 4070 SUPER) and takes about 2\,h.

\begin{table*}[t]\centering\footnotesize\setlength{\tabcolsep}{5.2pt}\renewcommand{\arraystretch}{1.0}
\caption{Copy-synthesis quality on the complete corpora. MUSHRA: mean score, 0--100 (Natural = hidden reference). Other columns: bold, best synthetic system, paired Wilcoxon $p<0.05$; underlined, best but not significant. FT: fine-tuned on the same 1\,h.}
\label{tab:t3_eval_combined}
\begin{tabular}{@{}ll c cc cccc c@{}}\toprule
 & & Subjective & \multicolumn{2}{c}{Signal} & \multicolumn{4}{c}{Predicted MOS $\uparrow$} & Content \\
\cmidrule(lr){3-3}\cmidrule(lr){4-5}\cmidrule(lr){6-9}\cmidrule(l){10-10}
Corpus & System & MUSHRA$\uparrow$ & MCD$\downarrow$ & LSD$\downarrow$ & UTMOS & UTMOSv2 & NISQA-TTS & SQUIM & CER\,(\%)$\downarrow$\,/\,SBS$\uparrow$ \\ \midrule
\multicolumn{10}{@{}l}{\textsc{Sentences}} \\[1pt]
CSMSC (zh) & \textcolor{gray}{\textit{Natural}} & \textcolor{gray}{97.5} & \textcolor{gray}{--} & \textcolor{gray}{--} & \textcolor{gray}{3.89} & \textcolor{gray}{3.25} & \textcolor{gray}{4.60} & \textcolor{gray}{4.38} & \textcolor{gray}{2.90} \\
\textcolor{gray}{\scriptsize 1{,}364 sent.} & ARIS (ours) & 87.5 & 3.40 & \textbf{7.38} & 3.01 & 2.50 & 4.47 & 4.25 & 3.17 \\
 & WORLD & 73.3 & \textbf{3.19} & 7.62 & 2.81 & 2.51 & 4.15 & 4.07 & \textbf{3.01} \\
 & KlattGrid & 7.0 & 23.50 & 46.86 & 1.34 & 1.79 & 3.03 & 2.51 & 7.05 \\
 & HiFi-Glot (FT) & 83.0 & 4.52 & 9.03 & \textbf{3.49} & \textbf{3.18} & \textbf{4.59} & \textbf{4.38} & 3.39 \\
\cmidrule(l){1-10}
HiFi-TTS (en) & \textcolor{gray}{\textit{Natural}} & \textcolor{gray}{96.8} & \textcolor{gray}{--} & \textcolor{gray}{--} & \textcolor{gray}{4.17} & \textcolor{gray}{3.46} & \textcolor{gray}{3.28} & \textcolor{gray}{4.40} & \textcolor{gray}{0.52} \\
\textcolor{gray}{\scriptsize 1{,}251 sent.} & ARIS (ours) & 62.2 & 3.72 & \textbf{7.25} & 3.66 & 2.97 & 2.91 & 4.33 & \underline{0.59} \\
 & WORLD & 68.2 & \textbf{3.66} & 7.84 & 3.60 & 2.57 & 3.02 & 4.21 & 0.61 \\
 & KlattGrid & 3.6 & 20.87 & 35.25 & 1.35 & 1.76 & 2.28 & 2.45 & 19.28 \\
 & HiFi-Glot (FT) & 70.5 & 4.93 & 8.77 & \textbf{3.88} & \textbf{3.34} & \textbf{3.12} & \textbf{4.40} & 0.91 \\
\midrule[0.8pt]
\multicolumn{10}{@{}l}{\textsc{Isolated syllables and words}} \\[1pt]
F024 (zh) & \textcolor{gray}{\textit{Natural}} & \textcolor{gray}{--} & \textcolor{gray}{--} & \textcolor{gray}{--} & \textcolor{gray}{3.13} & \textcolor{gray}{2.41} & \textcolor{gray}{3.67} & \textcolor{gray}{3.74} & \textcolor{gray}{--} \\
\textcolor{gray}{\scriptsize 1{,}517 syll.} & ARIS (ours) & -- & \textbf{3.45} & \textbf{7.20} & \textbf{2.70} & \textbf{1.93} & \textbf{3.39} & \textbf{3.60} & 0.923 \\
 & WORLD & -- & 3.82 & 8.22 & 2.37 & 1.83 & 3.10 & 3.11 & \textbf{0.925} \\
\cmidrule(l){1-10}
MALD (en) & \textcolor{gray}{\textit{Natural}} & \textcolor{gray}{--} & \textcolor{gray}{--} & \textcolor{gray}{--} & \textcolor{gray}{3.81} & \textcolor{gray}{2.91} & \textcolor{gray}{3.08} & \textcolor{gray}{4.22} & \textcolor{gray}{--} \\
\textcolor{gray}{\scriptsize 6{,}195 words} & ARIS (ours) & -- & 3.83 & 7.66 & 2.75 & 2.47 & 2.69 & 4.07 & 0.868 \\
 & WORLD & -- & \textbf{3.29} & \textbf{7.60} & \textbf{3.42} & \textbf{2.55} & \textbf{2.96} & \textbf{4.24} & \textbf{0.919} \\
\cmidrule(l){1-10}
BALDEY (nl) & \textcolor{gray}{\textit{Natural}} & \textcolor{gray}{--} & \textcolor{gray}{--} & \textcolor{gray}{--} & \textcolor{gray}{3.29} & \textcolor{gray}{3.03} & \textcolor{gray}{3.22} & \textcolor{gray}{4.19} & \textcolor{gray}{--} \\
\textcolor{gray}{\scriptsize 5{,}541 words} & ARIS (ours) & -- & 4.02 & \textbf{7.77} & 2.57 & \textbf{2.46} & 2.87 & \textbf{4.09} & 0.884 \\
 & WORLD & -- & \textbf{3.33} & 7.90 & \textbf{2.78} & 2.29 & \textbf{3.12} & 4.06 & \textbf{0.904} \\
\bottomrule\end{tabular}\end{table*}

\section{Experimental Setup}
\label{sec:setup}

\subsection{Data}
We train one model per single-speaker corpus, with an 8:1:1 split into training, validation and test sets. The Mandarin corpora are F024, a BLCU-SAIT speaker~\cite{blcusait} whose 1{,}517 monosyllables cover all legal syllables in the four tones (about 0.6\,h, 16\,kHz), and 1\,h of CSMSC~\cite{csmsc} sentences, downsampled from 48 to 24\,kHz. The other three corpora contain 1\,h each of English sentences (HiFi-TTS speaker 92~\cite{bakhturina2021hifitts}), English words (MALD~\cite{tucker2019mald}) and Dutch words (BALDEY~\cite{ernestus2015baldey}), all at 16\,kHz.

\subsection{Baselines}
The classic baselines are WORLD analysis--resynthesis~\cite{world} and Praat \emph{To KlattGrid (simple)}~\cite{weenink2009klattgrid}, which uses Praat's default analysis without noise sources and represents an off-the-shelf formant synthesizer. For manipulation, WORLD scales its $F_0$ (it cannot control single formants), while KlattGrid scales $F_0$ and $F_1$--$F_3$ through its pitch tier and \emph{Formula (frequencies)}. As a neural baseline on both sentence corpora, we fine-tune the released 1664\,h checkpoint of HiFi-Glot~\cite{hifiglot} with its official code for 10k additional steps on the same 1\,h that ARIS is trained on. Its 44.1\,kHz output is downsampled to the evaluation rate of each corpus, and for manipulation we scale its $F_0$ and $F_1$--$F_3$ inputs on voiced frames.

\subsection{Evaluation}
\emph{Naturalness}: since ARIS manipulates the recordings it is trained on, every system resynthesizes the complete corpora (all splits pooled). We score the output with MCD, LSD and the non-intrusive predictors UTMOS~\cite{utmos}, UTMOSv2~\cite{baba2024utmosv2}, NISQA-TTS~\cite{mittag2020nisqatts} and SQUIM~\cite{kumar2023squim}, computed with VERSA~\cite{shi2024versa} after normalization to $-26$\,dBFS, and measure content fidelity with Qwen3-ASR~\cite{shi2026qwen3asr} CER on sentences and SpeechBERTScore~\cite{saeki2024speechbertscore} on syllables and words. \emph{Controllability}: we scale $F_0$--$F_3$ of 60 sentences per corpus by 0.7--1.3, taking each system's own unmodified resynthesis times the factor as the target. Precision is then RMSE$_{95}$, the RMSE between the measured (RMVPE $F_0$, Praat Burg formants) and target tracks over the best 95\% of frames, which suppresses tracking jumps. \emph{Crosstalk}: for edits of 0.8--1.2, we measure the mean relative drift of the uncontrolled cues. On CSMSC we also edit $R_d$ (0.8--1.2) and noise gain ($\pm3$, $\pm6$\,dB) and score the edited audio as above. Differences are tested with paired Wilcoxon signed-rank tests ($p<0.05$).
Beyond these objective measures, we run only a small-scale MUSHRA test, because of budget constraints and because phonetic stimuli depend largely on each study's design. It covers copy-synthesized sentences from CSMSC and HiFi-TTS, rated against a hidden reference and a low-pass anchor. Word stimuli are omitted, since those from different systems were reported as hard to distinguish from each other. Ten phonetics or linguistics students, all native speakers of Mandarin or Cantonese, took part.

\begin{figure*}[t]\centering\includegraphics[width=\textwidth]{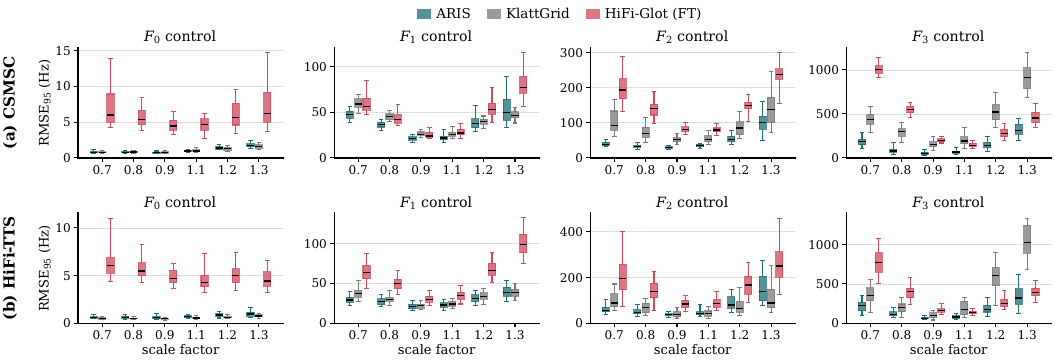}
\caption{Control precision: RMSE$_{95}$ of the controlled cue when $F_0$--$F_3$ are scaled by the given factor (60 sentences each) on (a) CSMSC and (b) HiFi-TTS. FT: HiFi-Glot fine-tuned on the same 1\,h. Boxes: quartiles; whiskers: 5th--95th percentiles.}\label{fig:precision}\end{figure*}

\section{Results}
\label{sec:results}

\subsection{Naturalness}
Table~\ref{tab:t3_eval_combined} summarizes copy-synthesis quality, with rankings unchanged on the test split. \emph{Mandarin sentences}: ARIS performs well at the signal level, with the lowest LSD of all systems (WORLD has a slightly lower MCD), and its NISQA-TTS and SQUIM scores come close to those of natural recordings. In MUSHRA it is also rated above both WORLD and HiFi-Glot (87.5 vs.\ 73.3 and 83.0). HiFi-Glot, in contrast, is weaker than ARIS and WORLD at the signal level, yet its predicted MOS approach the upper bound set by natural speech (NISQA-TTS 4.59 vs.\ 4.60). On inspection, however, HiFi-Glot does not reproduce the high-frequency content faithfully, adding spurious noise or harmonic components in that band, possibly because 1\,h of data and 10k steps are too little to adapt the pre-trained model. \emph{English sentences}: ARIS and WORLD perform at a similar level on the objective metrics, and HiFi-Glot again obtains the best predicted quality. It is also rated best in MUSHRA (70.5, vs.\ 68.2 for WORLD and 62.2 for ARIS), where scores are lower overall, possibly because the listeners are not native speakers of English. \emph{Words}: on Mandarin syllables ARIS outperforms WORLD on all signal and predicted-MOS metrics. On English words the order reverses, plausibly because the MALD speaker is male ($F_0$ 101\,Hz) while the formant priors of ARIS were set mainly for female voices. On Dutch words (female speaker), each system wins on some metrics. This pattern suggests that the simple (C)V(N) syllables and stable vowels of Mandarin favor the frame-wise vocal-tract model of ARIS. English and Dutch have more complex syllables and consonant clusters, and ARIS is less expressive for consonants such as nasals and laterals, where the spectral envelope of WORLD may cope better.

\begin{table}[t]\centering\footnotesize\setlength{\tabcolsep}{2.3pt}
\caption{ARIS crosstalk and edit quality on 60 CSMSC sentences: mean relative drift (\%) of each measured cue when the row parameter is edited (formants, $R_d$: 0.8--1.2; noise: $\pm3$, $\pm6$\,dB), and change in predicted MOS (v2: UTMOSv2; NISQA: NISQA-TTS) and CER (points) relative to unedited resynthesis. Bold: drift $\geq$ 1\%.}
\label{tab:crosstalk}
\begin{tabular}{@{}l cccc ccccc@{}}\toprule
 & \multicolumn{4}{c}{Drift (\%)} & \multicolumn{5}{c}{$\Delta$ quality} \\
\cmidrule(lr){2-5}\cmidrule(l){6-10}
Edited & $F_0$ & $F_1$ & $F_2$ & $F_3$ & UTMOS & v2 & NISQA & SQUIM & CER \\ \midrule
$F_0$  & --    & \textbf{1.7}  & 0.12 & 0.07 & $-$0.08 & $-$0.01 & $-$0.20 & $-$0.08 & $-$0.28 \\
$F_1$  & 0.01  & --    & 0.02 & 0.01 & $-$0.11 & $-$0.06 & $-$0.04 & $-$0.14 & $+$0.16 \\
$F_2$  & 0.00  & 0.04  & --   & 0.02 & $-$0.12 & $-$0.03 & $-$0.03 & $-$0.08 & $+$0.19 \\
$F_3$  & 0.00  & 0.02  & 0.04 & --   & $-$0.13 & $+$0.01 & $+$0.02 & $-$0.14 & $+$0.05 \\
$R_d$  & 0.00  & \textbf{1.6}  & 0.04 & 0.05 & $-$0.01 & $-$0.01 & 0.00 & $-$0.03 & $-$0.03 \\
Noise  & 0.01  & 0.07  & 0.01 & 0.02 & $-$0.07 & $-$0.02 & $-$0.07 & $-$0.09 & 0.00 \\
\bottomrule\end{tabular}\end{table}

\subsection{Controllability}
Fig.~\ref{fig:precision} shows that ARIS and KlattGrid scale $F_0$ with errors of about 1\,Hz, and HiFi-Glot with about 5\,Hz. Formant edits separate the systems more clearly: over all factors, the median $F_1$/$F_2$/$F_3$ errors of ARIS on CSMSC are 35.9/37.8/110.1\,Hz, compared with 41.3/72.3/346.9\,Hz for KlattGrid and 47.0/142.8/356.4\,Hz for HiFi-Glot. The ranking is the same on HiFi-TTS (Fig.~\ref{fig:precision}b). HiFi-Glot's $F_2$/$F_3$ errors are thus more than three times those of ARIS, possibly because its control mapping, learned from many speakers, fits the target speaker poorly. ARIS itself is weakest for large increases: it under-scales $F_1$ beyond 1.1$\times$, and $F_3{\times}1.3$ degrades all systems. Source and noise controls behave as intended: scaling $R_d$ by 0.7--1.3 changes $H_1-H_2$ monotonically from $-3.7$ to $+2.9$\,dB, close to the LF-model prediction ($-4.2$ to $+3.9$\,dB), and a $\pm6$\,dB noise gain changes unvoiced energy by $-5.9$/$+6.0$\,dB.

\subsection{Crosstalk}
We further analyze how editing one parameter affects the others. Because ARIS controls the DSP parameters directly, crosstalk is very small (Table~\ref{tab:crosstalk}): formant and noise edits move the other formants by at most 0.04\% and 0.07\%. The main exceptions are the effects of $F_0$ and $R_d$ edits on the measured $F_1$ (1.7\% and 1.6\%). These may partly be estimation errors caused by the edit, since both edits alter the low harmonics on which the $F_1$ estimate relies. Frame by frame (RMSE$_{95}$, with tracker noise), the ranking holds: the non-target formants drift by a median of 5.7/12.4/22.5\,Hz for ARIS/KlattGrid/HiFi-Glot on CSMSC. Editing also costs little quality: over 0.8--1.2$\times$, UTMOS falls by at most 0.13, against a 0.93 gap between natural and unedited audio, and CER stays within 2.6--3.1\%. Clear drops occur only at extreme factors ($-0.63$ UTMOS for $F_1\times0.7$). Overall, these results suggest that the controls of ARIS interact little under moderate edits.

\subsection{Ablation}\label{sec:ablation}
Table~\ref{tab:ablation} removes one component at a time. Without the formant loss $\mathcal{L}_F$, copy synthesis stays intact but control breaks on sentences: the explicit sections no longer track their formants, and the median $F_2$/$F_3$ errors rise six- to tenfold and drift fortyfold. It remains usable on syllables, which are dominated by steady vowels, so supervision matters most for the running speech common in phonetic experiments. Removing the residual poles has the opposite effect: the spectral distance degrades on both corpora, and the formant errors rise by 9--29\% because the explicit sections absorb detail beyond $F_1$--$F_3$.

\begin{table}[t]\centering\footnotesize\setlength{\tabcolsep}{3pt}
\caption{Ablations (50k steps, same data; test split). Control: median RMSE$_{95}$ (Hz) over 0.7--1.3 and mean drift (\%). Bold: best per corpus; $^\dagger$: differs from the full model ($p<0.05$).}
\label{tab:ablation}
\begin{tabular}{@{}l cc ccc@{}}\toprule
 & \multicolumn{2}{c}{Quality} & \multicolumn{3}{c}{Control} \\ \cmidrule(lr){2-3}\cmidrule(l){4-6}
Model & LSD$\downarrow$ & UTMOS$\uparrow$ & $F_2{\downarrow}$ & $F_3{\downarrow}$ & drift$\downarrow$ \\ \midrule
\multicolumn{6}{@{}l}{\emph{CSMSC 1\,h (sentences)}} \\
\quad full model & 7.47 & 3.04 & \textbf{37.8} & \textbf{110.1} & \textbf{0.02} \\
\quad w/o formant loss $\mathcal{L}_F$ & \textbf{7.44}$^\dagger$ & \textbf{3.14}$^\dagger$ & 359.6 & 658.7 & 0.80 \\
\quad w/o residual poles & 7.57$^\dagger$ & 3.09$^\dagger$ & 43.9 & 137.9 & 0.03 \\
\addlinespace[2pt]
\multicolumn{6}{@{}l}{\emph{F024 (syllables)}} \\
\quad full model & 7.37 & 2.73 & \textbf{43.6} & \textbf{93.1} & \textbf{0.10} \\
\quad w/o formant loss $\mathcal{L}_F$ & \textbf{7.31}$^\dagger$ & \textbf{2.87}$^\dagger$ & 48.3 & 107.9 & 0.13 \\
\quad w/o residual poles & 7.51$^\dagger$ & 2.72 & 50.7 & 120.3 & 0.11 \\
\bottomrule\end{tabular}\end{table}

\section{Conclusion}
\label{sec:conclusion}
We presented ARIS, a glass-box source--filter synthesizer that pairs a neural analysis with explicit DSP parameters and needs under 1\,h of one speaker. On five corpora it resynthesizes speech with quality comparable to WORLD, leads the Mandarin listening test, and edits single formants precisely with little effect on other cues, giving phoneticians a practical tool for constructing stimuli. HiFi-Glot attains higher predicted MOS, plausibly because adversarial training enhances high-frequency detail. Future work will add GAN- or flow-based post-enhancement and speaker-adaptive analytic poles for male and multi-speaker data.

\newpage
\section{Compliance with Ethical Standards}
All corpora are used under their respective licenses. Listening-test participants gave informed consent, and only ratings and anonymous IDs were stored.

\section{Acknowledgments}\label{ack}
We are grateful to Xin Wang (National Institute of Informatics, Japan) for many insightful discussions and for his generous encouragement throughout this work.

{\footnotesize\let\oldthebibliography\thebibliography
\renewcommand{\thebibliography}[1]{\oldthebibliography{#1}\setlength{\itemsep}{0pt}\setlength{\parskip}{0pt}}
\bibliographystyle{IEEEbib}
\bibliography{refs}}

@article{kawahara1999straight,
  author  = {Hideki Kawahara and Ikuyo Masuda-Katsuse and Alain de Cheveign{\'e}},
  title   = {Restructuring Speech Representations Using a Pitch-Adaptive Time--Frequency Smoothing and an Instantaneous-Frequency-Based {$F_0$} Extraction: Possible Role of a Repetitive Structure in Sounds},
  journal = {Speech Commun.},
  volume  = {27},
  number  = {3--4},
  pages   = {187--207},
  year    = {1999},
  doi     = {10.1016/S0167-6393(98)00085-5}
}

@inproceedings{wei2023rmvpe,
  author    = {Haojie Wei and others},
  title     = {{RMVPE}: A Robust Model for Vocal Pitch Estimation in Polyphonic Music},
  booktitle = {Proc. Interspeech},
  pages     = {5421--5425},
  year      = {2023},
  doi       = {10.21437/Interspeech.2023-528}
}

@inproceedings{malisz2019modern,
  author    = {Zofia Malisz and Gustav Eje Henter and Cassia Valentini-Botinhao and others},
  title     = {Modern Speech Synthesis for Phonetic Sciences: A Discussion and an Evaluation},
  booktitle = {Proc. ICPhS},
  pages     = {487--491},
  year      = {2019}
}

@article{liberman1957boundaries,
  author  = {Alvin M. Liberman and others},
  title   = {The Discrimination of Speech Sounds Within and Across Phoneme Boundaries},
  journal = {J. Exp. Psychol.},
  volume  = {54},
  number  = {5},
  pages   = {358--368},
  year    = {1957},
  doi      = {10.1037/h0044417}
}

@inproceedings{lisker1970voicing,
  author    = {Leigh Lisker and Arthur S. Abramson},
  title     = {The Voicing Dimension: Some Experiments in Comparative Phonetics},
  booktitle = {Proc. ICPhS},
  pages     = {563--567},
  year      = {1970}
}

@article{blicher1990mandarintone,
  author  = {D. L. Blicher and R. L. Diehl and L. B. Cohen},
  title   = {Effects of Syllable Duration on the Perception of the {Mandarin} Tone 2/Tone 3 Distinction: Evidence of Auditory Enhancement},
  journal = {J. Phonetics},
  volume  = {18},
  pages   = {37--49},
  year    = {1990},
  doi      = {10.1016/S0095-4470(19)30357-2}
}

@article{hillenbrand1999resynthesized,
  author  = {James M. Hillenbrand and Terrance M. Nearey},
  title   = {Identification of Resynthesized {/hVd/} Utterances: Effects of Formant Contour},
  journal = {J. Acoust. Soc. Am.},
  volume  = {105},
  number  = {6},
  pages   = {3509--3523},
  year    = {1999},
  doi     = {10.1121/1.424676}
}

@article{klatt,
  author  = {Dennis H. Klatt},
  title   = {Software for a Cascade/Parallel Formant Synthesizer},
  journal = {J. Acoust. Soc. Am.},
  volume  = {67},
  number  = {3},
  pages   = {971--995},
  year    = {1980},
  doi      = {10.1121/1.383940}
}

@article{klatt1990voicequality,
  author  = {Dennis H. Klatt and Laura C. Klatt},
  title   = {Analysis, Synthesis, and Perception of Voice Quality Variations Among Female and Male Talkers},
  journal = {J. Acoust. Soc. Am.},
  volume  = {87},
  number  = {2},
  pages   = {820--857},
  year    = {1990},
  doi      = {10.1121/1.398894}
}

@inproceedings{weenink2009klattgrid,
  author    = {David Weenink},
  title     = {The {KlattGrid} Speech Synthesizer},
  booktitle = {Proc. Interspeech},
  pages     = {2059--2062},
  year      = {2009},
  doi       = {10.21437/Interspeech.2009-591}
}

@inproceedings{wavebender,
  author    = {Gustavo Teodoro D{\"o}hler Beck and others},
  title     = {{Wavebender GAN}: An Architecture for Phonetically Meaningful Speech Manipulation},
  booktitle = {Proc. ICASSP},
  pages     = {6187--6191},
  year      = {2022},
  doi       = {10.1109/ICASSP43922.2022.9747442}
}

@article{schwaer2023multiscale,
  author  = {Simon Schw{\"a}r and Meinard M{\"u}ller},
  title   = {Multi-Scale Spectral Loss Revisited},
  journal = {IEEE Signal Process. Lett.},
  volume  = {30},
  pages   = {1712--1716},
  year    = {2023},
  doi     = {10.1109/LSP.2023.3333205}
}

@inproceedings{hifigan,
  author    = {Jungil Kong and Jaehyeon Kim and Jaekyoung Bae},
  title     = {{HiFi-GAN}: Generative Adversarial Networks for Efficient and High Fidelity Speech Synthesis},
  booktitle = {Proc. NeurIPS},
  volume    = {33},
  pages     = {17022--17033},
  year      = {2020},}

@inproceedings{golf,
  author    = {Chin-Yun Yu and Gy\"orgy Fazekas},
  title     = {Differentiable Time-Varying Linear Prediction in the Context of End-to-End Analysis-by-Synthesis},
  booktitle = {Proc. Interspeech},
  pages     = {1820--1824},
  year      = {2024},
  doi       = {10.21437/Interspeech.2024-1187},}

@article{yu2024golf,
  author  = {Chin-Yun Yu and Gy{\"o}rgy Fazekas},
  title   = {{GOLF}: A Singing Voice Synthesiser with Glottal Flow Wavetables and {LPC} Filters},
  journal = {Trans. ISMIR},
  volume  = {7},
  number  = {1},
  pages   = {316--330},
  year    = {2024},
  doi     = {10.5334/tismir.210}
}

@article{hifiglot,
  author  = {Yicheng Gu and others},
  title   = {{HiFi-Glot}: High-Fidelity Neural Formant Synthesis with Differentiable Resonant Filters},
  journal = {arXiv:2409.14823},
  year    = {2024},
  doi     = {10.48550/arXiv.2409.14823},}

@inproceedings{ddsp,
  author    = {Jesse Engel and others},
  title     = {{DDSP}: Differentiable Digital Signal Processing},
  booktitle = {Proc. ICLR},
  year      = {2020},}

@inproceedings{nsf,
  author    = {Xin Wang and Shinji Takaki and Junichi Yamagishi},
  title     = {Neural Source-Filter-Based Waveform Model for Statistical Parametric Speech Synthesis},
  booktitle = {Proc. ICASSP},
  pages     = {5916--5920},
  year      = {2019},
  doi      = {10.1109/ICASSP.2019.8682298}
}

@article{lf,
  author  = {Gunnar Fant and Johan Liljencrants and Qi-guang Lin},
  title   = {A Four-Parameter Model of Glottal Flow},
  journal = {STL-QPSR},
  volume  = {26},
  number  = {4},
  pages   = {1--13},
  year    = {1985}
}

@article{world,
  author  = {Masanori Morise and Fumiya Yokomori and Kenji Ozawa},
  title   = {{WORLD}: A Vocoder-Based High-Quality Speech Synthesis System for Real-Time Applications},
  journal = {IEICE Trans. Inf. Syst.},
  volume  = {E99-D},
  number  = {7},
  pages   = {1877--1884},
  year    = {2016},
  doi      = {10.1587/transinf.2015EDP7457}
}

@inproceedings{utmos,
  author    = {Takaaki Saeki and Detai Xin and Wataru Nakata and others},
  title     = {{UTMOS}: {UTokyo-SaruLab} System for {VoiceMOS} Challenge 2022},
  booktitle = {Proc. Interspeech},
  year      = {2022},
  doi      = {10.21437/Interspeech.2022-439}
}

@inproceedings{sawsing,
  author    = {Da-Yi Wu and Wen-Yi Hsiao and Fu-Rong Yang and others},
  title     = {{DDSP}-based Singing Vocoders: A New Subtractive-based Synthesizer and A Comprehensive Evaluation},
  booktitle = {Proc. ISMIR},
  year      = {2022},
  doi       = {10.5281/zenodo.7316600}
}

@misc{torchlpc,
  author       = {Chin-Yun Yu and Gy{\"o}rgy Fazekas},
  title        = {{torchlpc}: Fast and Differentiable Time-Domain All-Pole Filtering in {PyTorch}},
  howpublished = {\url{https://github.com/DiffAPF/torchlpc}},
  year         = {2025},}

@misc{csmsc,
  author       = {{DataBaker}},
  title        = {{Chinese Standard Mandarin Speech Corpus}},
  howpublished = {\url{https://www.data-baker.com/en/datasets/freeDatasets/}},
  year         = {2022}
}

@inproceedings{shi2024versa,
  title={{VERSA}: A Versatile Evaluation Toolkit for Speech, Audio, and Music},
  author={Shi, Jiatong and others},
  booktitle={Proc. NAACL (System Demonstrations)},
  year={2025}
}

@inproceedings{bakhturina2021hifitts,
  author    = {Evelina Bakhturina and others},
  title     = {{Hi-Fi} Multi-Speaker {English} {TTS} Dataset},
  booktitle = {Proc. Interspeech},
  pages     = {2776--2780},
  year      = {2021},
  doi       = {10.21437/Interspeech.2021-1599}
}

@article{tucker2019mald,
  author  = {Benjamin V. Tucker and Daniel Brenner and D. Kyle Danielson and others},
  title   = {The Massive Auditory Lexical Decision ({MALD}) database},
  journal = {Behav. Res. Methods},
  volume  = {51},
  number  = {3},
  pages   = {1187--1204},
  year    = {2019},
  doi     = {10.3758/s13428-018-1056-1}
}

@article{ernestus2015baldey,
  author  = {Mirjam Ernestus and Anne Cutler},
  title   = {{BALDEY}: A database of auditory lexical decisions},
  journal = {Q. J. Exp. Psychol.},
  volume  = {68},
  number  = {8},
  pages   = {1469--1488},
  year    = {2015},
  doi     = {10.1080/17470218.2014.984730}
}

@inproceedings{kumar2023squim,
  author    = {Anurag Kumar and Ke Tan and Zhaoheng Ni and others},
  title     = {{TorchAudio-Squim}: Reference-Less Speech Quality and Intelligibility Measures in {TorchAudio}},
  booktitle = {Proc. ICASSP},
  year      = {2023},
  doi       = {10.1109/ICASSP49357.2023.10096680}
}

@article{shi2026qwen3asr,
  author  = {Xian Shi and Xiong Wang and Zhifang Guo and others},
  title   = {{Qwen3-ASR} Technical Report},
  journal = {arXiv:2601.21337},
  year    = {2026}
}

@inproceedings{saeki2024speechbertscore,
  author    = {Takaaki Saeki and Soumi Maiti and Shinnosuke Takamichi and others},
  title     = {{SpeechBERTScore}: Reference-Aware Automatic Evaluation of Speech Generation Leveraging {NLP} Evaluation Metrics},
  booktitle = {Proc. Interspeech},
  pages     = {4943--4947},
  year      = {2024},
  doi       = {10.21437/Interspeech.2024-1508}
}

@misc{winn2019continuum,
  author       = {Matthew B. Winn},
  title        = {Make Formant Continuum: A {Praat} script for creating vowel formant continua from modified natural speech (version 38)},
  year         = {2019},
  howpublished = {\url{http://www.mattwinn.com/praat.html}}
}

@article{yoneyama2023usfgan,
  author  = {Yoneyama, Reo and Wu, Yi-Chiao and Toda, Tomoki},
  title   = {High-Fidelity and Pitch-Controllable Neural Vocoder Based on Unified Source-Filter Networks},
  journal = {IEEE/ACM Trans. Audio, Speech, Lang. Process.},
  volume  = {31},
  year    = {2023}
}

@inproceedings{yoneyama2022sifigan,
  author    = {Yoneyama, Reo and Wu, Yi-Chiao and Toda, Tomoki},
  title     = {Source-Filter {HiFi-GAN}: Fast and Pitch Controllable High-Fidelity Neural Vocoder},
  booktitle = {Proc. ICASSP},
  year      = {2023}
}

@inproceedings{mittag2020nisqatts,
  title={Deep learning based assessment of synthetic speech naturalness},
  author={Mittag, Gabriel and M{\"o}ller, Sebastian},
  booktitle={Proc. Interspeech},
  pages={1748--1752},
  year={2020}
}

@inproceedings{baba2024utmosv2,
  title={The T05 system for the {VoiceMOS} Challenge 2024: Transfer learning from deep image classifier to naturalness {MOS} prediction of high-quality synthetic speech},
  author={Baba, Kaito and others},
  booktitle={Proc. SLT},
  year={2024}
}

@misc{blcusait,
  author       = {{Beijing Language and Culture University}},
  title        = {{BLCU-SAIT}: Multi-modal {Chinese} Interlanguage Speech Corpus for Intelligent Pronunciation Teaching},
  howpublished = {\url{https://yuyanziyuan.blcu.edu.cn/en/info/1050/1296.htm}},
  note         = {Accessed Sep. 2026},}

@inproceedings{lameris2025voicequalityvc,
  author    = {Harm Lameris and Joakim Gustafsson and {\'E}va Sz{\'e}kely},
  title     = {{VoiceQualityVC}: A Voice Conversion System for Studying the Perceptual Effects of Voice Quality in Speech},
  booktitle = {Proc. Interspeech},
  year      = {2025},}

@article{baek2026dynamic,
  author  = {Seung-Cheol Baek and Seung-Goo Kim and Burkhard Maess and others},
  title   = {Dynamic acoustic-to-categorical representations of phonemes and prosody along ventral and dorsal speech streams},
  journal = {Nat. Commun.},
  volume  = {17},
  note    = {Art. no. 6082},
  year    = {2026},
  doi     = {10.1038/s41467-026-75240-0}}

@article{normanhaignere2025temporal,
  author  = {Sam V. Norman-Haignere and Menoua Keshishian and Orrin Devinsky and others},
  title   = {Temporal integration in human auditory cortex is predominantly yoked to absolute time},
  journal = {Nat. Neurosci.},
  volume  = {28},
  number  = {11},
  pages   = {2356--2365},
  year    = {2025},
  doi     = {10.1038/s41593-025-02060-8}}

@article{li2025continuum,
  author  = {Zhu Li and Yuqing Zhang and Yanlu Xie},
  title   = {Speech stimulus continuum synthesis using deep learning methods},
  journal = {Speech Commun.},
  volume  = {173},
  note    = {Art. no. 103266},
  year    = {2025},
  doi     = {10.1016/j.specom.2025.103266}}
\label{docend}
\end{document}